\documentstyle[12pt]{article}
\newcommand{\be}{\begin{equation}}
\newcommand{\ee}{\end{equation}}
\date{}

\def \gta {\mathrel{\vcenter
     {\hbox{$>$}\nointerlineskip\hbox{$\sim$}}}}

\begin{document}
\begin{titlepage}
\begin{flushright}
HD--THEP--93--25\\
CRN 93--34\\
Saclay T93/076\\
July 1993
\end{flushright}
\vspace{1cm}
\begin{center}
{\bf\LARGE Particle Spectrum in Supersymmetric Models }\\
\vspace{.3cm}
{\bf\LARGE with a Gauge Singlet}\\
\vspace{.8cm}
Ulrich Ellwanger\footnote{Supported by  a DFG Heisenberg fellowship,
e-mail: I96 at VM.URZ.UNI-HEIDELBERG.DE}\\
\vspace{.3cm}
Institut  f\"ur Theoretische Physik\\
Universit\"at Heidelberg\\
Philosophenweg 16, D-69120 Heidelberg, FRG\\
\vspace{.8cm}
Michel Rausch de Traubenberg\footnote{
e-mail: G092MR at FRCCSC21.BITNET}\\
\vspace{.3cm}
Physique Th\'eorique\\
Centre de Recherches Nucl\'eaires (CRN)\\
B.P. 20 CRO, F-67037 Strasbourg Cedex, France\\
\vspace{.8cm}
Carlos A. Savoy\footnote{
e-mail: SAVOY at AMOCO.SACLAY.CEA.FR}\\
\vspace{.3cm}
Service de Physique Th\'eorique, CEA-Saclay\\
F-91191 Gif-sur-Yvette Cedex, France\\
\vspace{1cm}
\end{center}

{\bf Abstract:}\\
\parbox[t]{\textwidth}{ We scan the complete parameter space of the
supersymmetric standard model extended by a gauge singlet, which is
compatible with the following constraints: universal soft supersymmetry
breaking terms at the GUT scale, finite running Yukawa couplings up to
the GUT scale and present experimental bounds on all sparticles, Higgs
scalar and top quark. The full radiative corrections to the Higgs
potential due to the top/stop sector are included. We find a lower
limit on the gluino mass of 160 GeV,
upper limits on the lightest neutral scalar Higgs mass dependent on
$m_{top}$ and the size of the soft supersymmetry breaking terms, and
the possibility of a Higgs scalar as light as 10 GeV,
but with reduced couplings to the Z boson.}

\end{titlepage}
\newpage
\setcounter{section}{1}
The main task of present and future accelerators is the search for
the Higgs boson, the top quark and, if existent, the new
particles predicted by supersymmetry. Predictions on the particle
spectrum in supersymmetric models can, however, only be made under
additional assumptions: One can invoke some conditions
on the absence of fine tuning, one can assume the running Yukawa
couplings not to diverge below a GUT scale, and
one can make assumptions on the parameters
specifying the soft supersymmetry breaking terms (the gaugino masses,
the masses of the scalars and the trilinear couplings among the scalars).

Most of the recent
discussions of the allowed space of parameters and
the particle spectrum, under various of the above assumptions, have
appeared within the minimal supersymmetric extension of the standard
model (MSSM) \cite{1}-\cite{8}. Also most of the experimental analyses
have been done in the framework of the MSSM. There the Higgs
sector is specified by just two unknown parameters, which include a
supersymmetric mass term $\mu$ for the two Higgs doublets. This $\mu$ -
term is actually a nuisance, since an explanation of its presence and a
reason, why it should be of the same order of magnitude as the soft
susy breakings, requires additional assumptions like
radiative corrections involving a GUT sector \cite{9},
nonrenormalizable interactions within supergravity \cite{10}, \cite{11},
 nontrivial K\"ahler potentials \cite{12} or accidental
global symmetries in the superpotential \cite{13}.

It is important, however, to extend theoretical and experimental studies
beyond the MSSM and consider supersymmetric extensions such as the one
involving one gauge singlet ((M+1)SSM) \cite{14}-\cite{18}.
For instance, the upper bound
on the lightest Higgs boson mass within the MSSM is violated already
in this (M+1)SSM. Furthermore the $\mu$ - term
can be omitted, and the entire superpotential can be chosen to be
scale invariant, i.e. to include only dimensionless Yukawa couplings, by
imposing a discrete $Z_3$-symmetry.
This is also the typical structure emerging from superstring
motivated scenarios.

Under the assumption of a scale invariant superpotential and universal
soft susy breakings at the GUT scale the Higgs sector of the this model
involves five independent parameters, two Yukawa couplings in the
superpotential and three universal soft susy breakings
at the GUT scale. Due to the importance of the top/stop induced
radiative corrections (renormalization effects from the GUT scale down
to the electroweak scale as well as radiative corrections to the Higgs
potential) the top Yukawa coupling appears as a sixth important
parameter. On the other hand the knowledge of the $Z/W$ masses reduces
the number of unknown parameters again to five.

Some investigations of the parameter space of the (M+1)SSM have been
performed in \cite{16}. If universality at the GUT scale was assumed,
however,
the authors confined themselves to soft susy breaking triggered
by gaugino masses only. Recently upper bounds on the mass of the
lightest neutral Higgs scalar under the only assumption of
the absence of fine tuning \cite{19} or finite
running Yukawa couplings below $M_{GUT}$ have been the subject of
detailed investigations \cite{20}-\cite{30}.

A scan of the complete five dimensional
parameter space assuming
universal soft susy breakings at $M_{GUT}$ and hence $SU(2)\times U(1)$
symmetry breaking triggered by radiative
corrections, providing a stable vacuum
and respecting the present experimental constraints, is substantially
more involved and is the task of the present paper as well as a
forthcoming publication \cite{31}.
It is not even clear, a priori, whether all conditions can
be satisfied simultaneously,
and whether the previously derived bounds can be saturated
under these additional conditions.

Let us describe the model and our procedure. The relevant part of
the superpotential has the form
\be\label{1}
W=h_t Q\cdot H_2 T^c_R+\lambda H_1\cdot H_2 S+\frac{1}{3} \kappa S^3\ee

where colour indices are suppressed and
\be\label{2}
Q= {T_L \choose B_L}, H_1={H^0_1 \choose H^-_1}, H_2=
{H^+_2 \choose H^0_2},\quad Q\cdot H=Q_i\epsilon^{ij}H_j\ {\rm etc.}
\ee
The scalar
potential contains the standard $F$- and $D$-terms, the soft
susy breaking terms and in addition the one loop radiative correction
of the form
\be\label{3}
V^{rad}=\frac{1}{64\pi^2}{\rm Str}[M^4\ln(M^2/Q^2)].\ee

The soft supersymmetry breaking gaugino masses, trilinear couplings and
scalar masses are given by
\begin{eqnarray}\label{4}
&&(\mu_1\lambda_1\lambda_1+\mu_2\lambda_2\lambda_2+
\mu_3\lambda_3\lambda_3+
h_tA_tQ\cdot H_2T^c_R+\lambda  A_\lambda
H_1\cdot H_2S+\frac{1}{3}\kappa A_{\kappa} S^3)\ +\ {\rm h.c.}\nonumber\\
&&+m^2_1|H_1|^2+m^2_2|H_2|^2+m^2_S|S|^2+m^2_Q|Q|^2+m^2_T|T^c_R|^2+...
\end{eqnarray}
where $\lambda_1$, $\lambda_2$ and $\lambda_3$ denote the gauginos of the
$U(1)_Y$, $SU(2)$ and $SU(3)$ gauge groups respectively.

Concerning the radiative corrections (3)
we only take top quark and squark loops into account. The
corresponding bottom contributions are neglicible assuming $\tan\beta
 = \langle H_2\rangle / \langle H_1\rangle < 20$, and the
gauge sector is known to play no important role \cite{32}, \cite{33}.
This also holds
for the extended Higgs sector, since the Yukawa couplings $\lambda$ and
$\kappa$ will turn out to be small.
Indeed, we eventually find $\lambda^2 < g^2_2$, where $g_2$ is the
$SU(2)$ gauge coupling, a sufficient condition to forbid the breaking of
electromagnetism in the Higgs sector \cite{34}. It has also been shown
that supersymmetry prevents the spontaneous breaking of $CP$ \cite{35},
so that the vevs of the
fields $H_1,H_2$ and $S$ are of the form
\be\label{5}
\langle H_1\rangle={h_1 \choose 0},\quad \langle H_2\rangle
={0 \choose h_2},\quad \langle S\rangle
=s\ee
with $h_1$, $h_2$ and $s$ real.
The Higgs vev dependent top quark mass
appearing in (3) is given by
\be\label{6}
m_t=h_th_2\ee
and the top squark mass matrix, in the basis $(T_R^c,T_L^*)$, by
\be\label{7}
\left(\begin{array}{cc}
h_t^2h_2^2+m^2_T-\frac{g_1^2}{3}(h_2^2-h_1^2)&
h_t(A_t h_2+\lambda sh_1)\\
h_t(A_th_2+\lambda sh_1)&
h^2_th_2^2+m^2_Q+(\frac{g_1^2}{12}-\frac{g_2^2}{4})(h_2^2-h_1^2)
\end{array}
\right).\ee

The equations for extrema of the
full scalar potential in the directions (5) in field space read
\be\label{8a}
h_1[m^2_1+\lambda^2(h_2^2+s^2)+\frac{g^2_1+g_2^2}{4}(h^2_1-h^2_2)]+
\lambda
h_2s(\kappa s+A_\lambda)+\frac{1}{2}\partial V^{rad}/\partial h_1=0,\ee
\be\label{9}
h_2[m^2_2+\lambda^2(h_1^2+s^2)+\frac{g^2_1+g_2^2}{4}(h^2_2-h^2_1)]+
\lambda
h_1s(\kappa s+A_\lambda)+\frac{1}{2}\partial V^{rad}/\partial h_2 =0,\ee
\be\label{10}
s[m^2_S+\lambda^2(h_1^2+h^2_2)+2\kappa^2s^2+2\lambda \kappa
h_1h_2+\kappa A_{\kappa}s]+\lambda A_\lambda h_1h_2+\frac{1}{2}\partial
V^{rad}/\partial s =0.\ee

In order to implement the assumption of finite Yukawa couplings
up to $10^{16}$ GeV and universal soft susy breakings at $10^{16}$
GeV we need the renormalization group equations for $\lambda$, $\kappa$
and $h_t$ and the susy breakings $m_1^2$, $m_2^2$, $m_S^2$, $m_Q^2$,
$m_T^2$, $A_\lambda$, $A_\kappa$ and $A_t$. These can be found in
\cite{15}, \cite{36}.

We scan the parameter space of the model as follows:
First we fix $\sim 3000$ different combinations of the Yukawa
couplings $\lambda_0$, $\kappa_0$ and $h_{t0}$ at the GUT scale in the range
from
0 to 3. In each case we integrate the renormalization group equations
numerically down to the electroweak scale of $O(100)$ GeV. Thereby
we determine the "low energy" Yukawa couplings $\lambda$, $\kappa$ and
$h_t$ as well as the coefficients of the expansion of the "low energy"
susy breakings $m_1^2$, ..., $A_t$ in terms of the three bare susy
breakings $\mu_0$, $A_0$ and $m_0^2$:
\begin{eqnarray}\label{11}
m_i^2&=&a_i\mu_0^2+b_iA_0^2+c_i\mu_0A_0+d_im_0^2,\quad i=1,2,S,Q,T;
\nonumber\\
A_a&=&e_aA_0+f_a\mu_0,\quad a=t,\lambda,\kappa.
\end{eqnarray}

The coefficients $a_i$, ..., $f_a$ are completely specified by the
three Yukawa couplings and the gauge couplings and are computed
numerically. Next, in each case,
we scan over $\sim 5000$ different values of the singlet vev $s$ and
$\tan\beta$ with
\begin{eqnarray}\label{12}
0 < s < 10\quad TeV,\nonumber\\
1 < |\tan\beta| < 20.
\end{eqnarray}
In each of those $\sim 15\times 10^{6}$ cases, we insert the specified
values for all three vevs $h_1$, $h_2$ and $s$ together with the
expansions (11) into the minimization equations (8 - 10). These allow
us to compute the three high energy susy breakings $\mu_0$, $m_0^2$ and
$A_0$.

Having fixed all parameters of the low energy potential we next
check, whether the present choice of vevs corresponds to an absolute
minimum of the potential. Thus we compute the energy density and
compare it with the extrema of the potential where either
$h_1$, $h_2$ or $s$ vanishes.
Large values of the trilinear couplings can induce squark or slepton
vevs, which would break color and/or electromagnetism \cite{37},
\cite{15}, \cite{38}.
The most dangerous possibility turns out to be $U(1)_{e.m.}$ breaking
by a selectron vev, against which we check after having computed the
required trilinear coupling and slepton masses and imposing the
condition
\be\label{13}
A_E^2 < 3(m_E^2+m_L^2+m_1^2)
\ee
In the remaining cases we proceed to calculate the physical masses of
the particles (their expressions in terms of the low energy parameters
are given in the literature \cite {16}, \cite {17}, \cite {19}).
In addition we compute the masses of the selectrons, sneutrino,
top quark, top squarks, gluino as well as the coupling of the lightest
neutral Higgs scalar to the Z boson. Then we impose the following
experimental constraints: We demand the sparticles which could be
pair produced in $Z_0$ decay (selectrons, sneutrino, squarks, charged
Higgs scalar and charginos) to be heavier than 45 GeV, the gluino to be
heavier than 110 GeV, the top quark to be heavier than 110 GeV and the
lightest neutral Higgs scalar either to be heavier than 58 GeV or to
have a reduced coupling to the Z boson according to a recent analysis
of the LEP experiments, which is general enough to include extensions
of the MSSM as the present one \cite{39}.

At the end we are still left with a considerable range of parameters and
masses, which satisfy all the constraints. The complete range of
Yukawa couplings, soft susy breakings, $s$ and $\tan\beta$ found by
our procedure is given by
\begin{eqnarray}\label{14}
0<&\lambda_0&<.55\nonumber\\
0<&\kappa_0&<.65\nonumber\\
.25<&h_{t0}&<3\nonumber\\
60\quad GeV<&|\mu_0|&<4\quad TeV\nonumber\\
0<&m_0&<3.5\quad TeV\nonumber\\
25\quad GeV<&|A_0|&<10\quad TeV\nonumber\\
800\quad GeV<&s&<10\quad TeV\nonumber\\
2<&|\tan\beta|&<20
.\end{eqnarray}

The lower limit on $h_{t0}$ is a reflection of our lower bound on
$m_{top}$
of 110 GeV. The nonvanishing lower limit on $A_0$ indicates that a
scenario with gaugino masses as the only source of susy breaking is not
allowed. (This limit is due to the present experimental
constraints on the lightest neutral Higgs scalar.)
A striking feature of these solutions is the large vev of the singlet
$s$ with respect to the weak interaction scale. As a matter of fact
the upper limits
on the susy breakings are not genuine, but a
reflection of our upper limit (12) on $s$.
At low energies the Yukawa couplings vary, as implied by (14), between
0 and .4 in the case of $\lambda$ and $\kappa$, and between .7 and 1.1
in the case of $h_t$.

In view of the large values of $s$ that characterise the solutions,
it is possible to understand some features of the results in terms of
approximate analytic solutions of the minimization equations (8) - (10)
and the RG equations. In the limit $s^2>>h_1^2+h_2^2=(174 GeV)^2$ the
stable
solutions of (10) are given by the relations
\begin{eqnarray}\label{15}
4\kappa s &=& -A_{\kappa}+\sqrt{A_{\kappa}^2-8m_S^2}\nonumber\\
A_{\kappa}^2 &>& 8m_S^2.
\end{eqnarray}
As a matter of fact, with $s$ given by (15), eqs. (8) and (9) are close
to the minimum conditions in the MSSM with the usual parameters $B$
and $\mu$ given by
\be\label{16}
B=A_{\lambda}-\kappa s,\qquad \mu=\lambda s.
\ee
It has to be noticed, however, that here the spectrum is richer and the
deviations from the MSSM physically relevant. Only in the mathematical
limit $s\to\infty$; $\lambda,\kappa\to 0$ with $\lambda s$, $\kappa s$
fixed the MSSM is reproduced.

To proceed let us make another approximation and neglect terms of
\newline $O(\kappa^2/h_t^2,\lambda^2/h_t^2)$ in the RG equations of
the Yukawa
couplings and soft breaking terms. From the solutions of the RGEs
analogous to those given in \cite{40} one can easily infer the low energy
Yukawa couplings ond soft terms to $O(\kappa^2/h_t^2,\lambda^2/h_t^2)$
in terms of the high energy ones. They turn out to be as follows:
\begin{eqnarray}\label{17}
\kappa^2&\simeq&\kappa_0^2\nonumber\\
\lambda^2&\simeq&1.86\lambda_0^2(1-.822 h_t^2)^{1/2} \nonumber\\
A_{\lambda}&\simeq&A_0(1-.411 h_t^2)-\mu_0(.59-.86 h_t^2)\nonumber\\
A_{\kappa}&\simeq&A_0\nonumber\\
m_S^2&\simeq&m_0^2\nonumber\\
m_1^2&\simeq&m_0^2+.53\mu_0^2\nonumber\\
m_2^2&\simeq&m_0^2(1-1.644 h_t^2)+.53\mu_0^2-3h_t^2[.137A_0^2
(1-.822h_t^2)\nonumber\\
&&+.56A_0\mu_0(1-.822h_t^2)+\mu_0^2(1.65-.50h_t^2)]
\end{eqnarray}
(see \cite{31} for details). Thus in the small $\kappa$, $\lambda$
approximation the condition on $A_{\kappa}$ in (15) gives
\be\label{18}
A_0^2 > 8 m_0^2.
\ee
At this stage we must check against colour and/or electromagnetism
breaking solutions. From expressions analogous to (17) we find the
parameters of (13) calculated at a scale of $O(A_E/h_e)\sim 10^8$ GeV,
hence the condition (13) can be translated into the following condition
on the parameters at the GUT scale in the limit of small $\kappa$,
$\lambda$:
\be\label{19}
A_0 < .4\mu_0+\sqrt{9m_0^2+2\mu_0}
\ee
The approximations involved in (17) - (19) improve as
$\lambda,\kappa\to 0$ with $\lambda s$, $\kappa s$ fixed. Therefore, the
(M+1)SSM includes the MSSM as a limiting case, but with a further
restriction on the parameters as given by these relations, at least
under the assumption of universality of the soft terms at the
unification scale. But even if this assumption is given up, eq.(15),
that follows from (10) in this limit, provides a restriction in the
resulting MSSM, whatever the boundary conditions at high energies may
be.

For larger values of $\kappa$, $\lambda$ the corresponding values for
the soft parameters increase and the overall solutions may require
some degree of finetuning of $A_0$, $\mu_0$ and $m_0^2$. We looked for
correlations among these parameters (at fixed Yukawa couplings), but
found only a relatively weak one among $m_0$ and $A_0$ of the
form
\be\label{20}
A_0\sim3m_0\pm 1\quad TeV.\ee

The essential features of the particle spectrum are as follows: The
top quark mass is bounded from above by 190 GeV, which is a well known
consequence of a finite value of $h_{t0}$. There are no obvious
correlations between the top quark mass and the scale of susy breaking.

For the mass of the lightest neutral Higgs scalar we obtain an upper
bound depending on $m_{top}$ and the soft susy breakings $A_t$, $m_Q$
and $m_T$. Here the complete radiative corrections induced by $V^{rad}$
of eq. (4) are seen to play an important role: For $m_{top}\simeq 190$
GeV and $A_t$, $m_Q$ and $m_T$ $\gta$ 2 TeV we have cases where the
tree level part of the Higgs potential would suggest a mass of the
lightest neutral Higgs scalar of $\sim M_Z$, the effect of the leading
logarithms of the form $\sim \ln{(m^2_{stop}/m^2_{top})}$ increase it to
$\sim 145$ GeV, and the remaining corrections to the Higgs potential
as computed in \cite{27}-\cite{30} increase it to $\sim 155$ GeV.
This value
constitutes an upper bound on $M_H$ for the range (14) of the soft
susy breakings; if, instead, we assume upper limits on $A_t$ or
$m_{stop}$ of $\sim 1$ TeV, this upper bound decreases to $\sim 140$
GeV. In fig.(1) we plot the bound on $M_H$, for the different ranges
of the soft susy breakings, versus $m_{top}$. We also plot the bound
based solely on the assumption of perturbative Yukawa couplings below
$M_{GUT}$ within this model (again with upper limits on $A_t$ or
$m_{stop}$ of $\sim 1$ TeV, taken from \cite{28}). This shows the
decrease
of the upper bound, mainly at moderate values of $m_{top}$, due to the
additional assumption of universal soft susy breakings. It is related to
the fact that we found an upper limit on the Yukawa
coupling $\lambda_0$ (see eq. (14)).

In the case of light neutral Higgs scalars we find the interesting
possibility that this particle could contain a large admixture of the
gauge singlet field $S$, and hence have a substantially reduced or
even vanishing coupling to the Z boson. (This possibility has also been
observed in \cite{16}, \cite{26}.) It allows Higgs masses as light as
10 GeV to be
compatible with up to date unsuccessful Higgs searches. This feature
persists up to Higgs masses of $\sim 100$ GeV, whereas for Higgs masses
beyond 125 GeV its coupling to the Z boson has to be very close to the
one of the standard model Higgs boson.

Concerning the charged Higgs scalar, charginos, stops and sleptons we
find that each of them could be as light as 45 GeV and thus be
detectable in the near future. However, the combination of the
experimental lower bounds on these particle masses results in the
lower limit of $\sim 60$ GeV on the bare susy breaking gaugino mass
term $\mu_0$, which implies a lower bound on the gluino mass of
$\sim 160$ GeV. The finding of a lighter gluino would thus imply a
violation of our underlying assumption of universality.

As is clear from the possible range of susy breakings, all new
particles implied by supersymmetry are also allowed to be very heavy;
upper limits on their masses can be obtained invoking some
condition on the absence of fine tuning, which is not the purpose of
the present paper.

Our procedure allows us to study
a multitude of correlations among these masses.
In fig.(2) we plot the allowed range of the masses of the lightest
chargino and the lightest top squark versus the gluino mass. This plot
explicitly shows the common increase of the sparticle
masses, but also how difficult it is to deduce one mass from the other
below 1 TeV under our general assumptions. More correlations will be
discussed in a separate more extended paper \cite{31}.

In conclusion, we have shown that the supersymmetric extension of the
standard model involving a gauge singlet, and assuming universal susy
breakings at the GUT scale, can well satisfy all present experimental
constraints. A scenario with the gaugino mass as the only seed of soft
susy breaking at $10^{16}$ GeV is, however, excluded. A gluino below 160
GeV is not allowed within this model. For moderate soft susy
breakings (below 1 TeV) the upper bound on the lightest neutral Higgs
scalar varies between 100 GeV for $m_{top}=100$ GeV and 140 GeV for
$m_{top}=190$ GeV. The possibility of a light Higgs scalar with reduced
couplings to the Z boson leads to interesting tasks for experiments.
Clearly our procedure will allow us to make more refined
predictions as soon as more experimental information (e.g. on $m_{top}$)
is available, or if theoretical assumptions on the soft susy breakings
or on the Yukawa couplings are made.

\section*{Figure Captions}
\begin{description}
\item{Fig. 1:} Upper bound on the mass of the lightest neutral Higgs
scalar versus $m_{top}$. Full line: result for the range (18) of
the soft susy breakings, dashed line: result with upper limits on $A_t$
or $m_{stop}$ of $\sim 1$ TeV, dotted line: result with the only
assumption of perturbative Yukawa couplings below
$M_{GUT}$ within this model (again with upper limits on $A_t$ or
$m_{stop}$ of $\sim 1$ TeV, taken from \cite{28}).
\item{Fig. 2:} Allowed range of the masses of the lightest
chargino (within the full lines) and the lightest top squark
 (within the dashed lines) versus the gluino mass.
\end{description}

\end{document}